# A new formulation of asset trading games in continuous time with essential forcing of variation exponent

KEI TAKEUCHI[1], MASAYUKI KUMON[2] and AKIMICHI TAKEMURA[3]

[1]*Emeritus, Graduate School of Economics, University of Tokyo, 7-3-1 Hongo, Bunkyo-ku, Tokyo 113-8656, Japan.* E-mail: *kei.takeuchi@wind.ocn.ne.jp*
[2]E-mail: *masayuki_kumon@smile.odn.ne.jp*
[3]*Graduate School of Information Science and Technology, University of Tokyo, 7-3-1 Hongo, Bunkyo-ku, Tokyo 113-8656, Japan.* E-mail: *takemura@stat.t.u-tokyo.ac.jp*

We introduce a new formulation of asset trading games in continuous time in the framework of the game-theoretic probability established by Shafer and Vovk (*Probability and Finance: It's Only a Game!* (2001) Wiley). In our formulation, the market moves continuously, but an investor trades in discrete times, which can depend on the past path of the market. We prove that an investor can essentially force that the asset price path behaves with the variation exponent exactly equal to two. Our proof is based on embedding high-frequency discrete-time games into the continuous-time game and the use of the Bayesian strategy of Kumon, Takemura and Takeuchi (*Stoch. Anal. Appl.* **26** (2008) 1161–1180) for discrete-time coin-tossing games. We also show that the main growth part of the investor's capital processes is clearly described by the information quantities, which are derived from the Kullback–Leibler information with respect to the empirical fluctuation of the asset price.

*Keywords:* Bayesian strategy; beta-binomial distribution; game-theoretic probability; Hölder exponent; Kullback–Leibler information; modulus of continuity; square root of *dt* effect

## 1. Introduction

In this paper, we present a new formulation of asset trading games in continuous time, in the framework of the game-theoretic probability of Shafer and Vovk [15]. In the book by Shafer and Vovk, continuous-time games are formulated as limits of discrete-time games by using techniques of nonstandard analysis. Although their approach is rigorously formulated in the framework of nonstandard analysis, we give another formulation of continuous-time games in the game-theoretic probability, which is tractable within the conventional theory of analysis.

An asset trading game is a complete information game between an investor and the market. Following Chapter 9 of Shafer and Vovk [15], we denote these two players as "Investor" and "Market". In our formulation, Market moves continuously, but Investor moves in discrete times, depending on the past path of Market. The trading times of Investor need not be equally spaced. In this paper, we mainly consider "limit order" strategy (rather than the "market order" strategy) of Investor. In the limit order strategy, Investor trades a financial asset when the asset price or the increment of the asset price hits a certain level. We shall prove that by a high-frequency limit order type Bayesian strategy, Investor can essentially force the variation exponent of two in the price path of Market. The precise definition of essential forcing will be given in Section 2.

In an infinitely repeated series of fair betting games, a gambler cannot make gain with certainty. This fact has been formulated and proven in the theory of martingales. But when the games are favorable to a gambler, for example, if the results of the games are stochastically independent with positive expected value, to what extent can he exploit the situation and what would be a good strategy to adopt? Several years after the advent of Shannon's celebrated work [16], this problem was first systematically studied by Kelly [9] in relation to the betting game interpretation of Shannon's mutual information quantity. In this spirit, betting games have been investigated by information theorists, which led to the notion of Cover's universal portfolios [2, 3]. One of the present authors also wrote a note on it about forty years ago in Japanese, presenting the results in [18].

Recently, Shafer and Vovk originated a new, attractive field of game-theoretic probability and finance [15]. The most important point concerning their approach is that stochastic behavior of Market is not assumed a priori, but follows from the protocol of the game between Investor and Market. Shafer and Vovk established the general fact that in order to prevent Investor from making an infinitely large gain, Market must behave as if he is stochastic and make the game fair in a stochastic sense. However, the question remains as to what Investor can make from Market's failure to do so. This issue was treated by Kumon and Takemura [10], where it is proved that when Market's moves are bounded, a simple strategy forces the strong law of large numbers (SLLN) with a convergence rate of $\mathrm{O}(\sqrt{\log n/n})$. Kumon, Takemura and Takeuchi [11] proved several versions of SLLN for the case where Market's moves are unbounded. For coin-tossing games, Kumon, Takemura and Takeuchi [12] considered a class of Bayesian strategies for Investor and established the important fact that if Market violates SLLN, then Investor can increase his capital exponentially fast and the exponential growth rate is precisely described in terms of the Kullback–Leibler information between the average of Market's moves when he violates SLLN and the average when he observes SLLN.

In this paper, we apply the results of [12] to asset trading games in continuous time. We consider implications of high-frequency limit order type Bayesian strategies and prove that Investor can make arbitrarily large gain if Market does not move jaggedly with the variation exponent equal to two. In the mathematical finance literature, this phenomenon has long been recognized and understood as the fact that fractional Brownian motion with the Hölder exponent $H \neq 1/2$ is not a semimartingale; see Rogers [14], Section 4.2 of Embrechts and Maejima [6] and Section 3 of Hobson [7]. Kunitomo [13] presented a similar result earlier. Note that in the measure-theoretic approach, these results require strong stochastic assumptions. Vovk and Shafer [23] treated the $\sqrt{dt}$ effect using nonstandard analysis. The game-theoretic approach in the present paper and in [23] is advantageous because no probabilistic model, such as fractional Brownian motion, is imposed on the paths of Market. It can be an arbitrary continuous path in our formulation. Another fundamental strength of the game-theoretic approach is that we can give statements on an individual path of Market, whereas in measure-theoretic probability, one can only make statements on measurable sets of the space of appropriate paths.

This paper is organized as follows. In Section 2, we formulate asset trading games and introduce the necessary notation and definitions. We also review the results on the Bayesian strategy of [12] for discrete-time games embedded into the continuous-time game. We investigate the consequences of high-frequency Bayesian strategy in Section 3 and establish that Market is essentially forced to move with the variation exponent exactly equal to two. We end the paper with some concluding remarks in Section 4.

## 2. Asset trading games in continuous time

In this section, we formulate asset trading games in continuous time and introduce appropriate notation and definitions. We begin with an informal description of asset trading games in continuous time and their embedded discrete-time games in Section 2.1. More precise definitions of the game and the move spaces of the players are presented in Section 2.2. In particular, we will define the notion of essential forcing of an event by Investor. In Section 2.3, we review notions of the variation exponent and the Hölder exponent. In Section 2.4, we summarize results on Bayesian strategy for coin-tossing games in [12].

### 2.1. Formulation of asset trading games in continuous time

Suppose that there is a financial asset which is traded in a market in continuous time. Let $S(t)$ denote the price of the unit amount of the asset at time $t$. We assume that $S(t)$ is positive and a continuous function of $t$. We consider the price path $S(\cdot)$ to be chosen by the player "Market". "Investor" enters the market at time $t = t_0 = 0$ (knowing the initial price $S(0)$) with the initial capital of $\mathcal{K}(0) = 1$ and he can buy or sell any amount of the asset at any time, provided that his capital always remains nonnegative. It is assumed that Investor can trade only at discrete time points $0 = t_0 < t_1 < t_2 < \cdots$, although he can decide the trading time $t_i$ and the amount he trades at $t_i$ based on the path of $S(t)$ up to time $t_i$. Since $S(t)$ is continuous, when we say "up to time $t_i$", we do not need to distinguish whether Investor is allowed to use the value $S(t_i)$ or not. His repeated tradings up to time $t_i$ also decide the amount $M_i$ of the asset he holds for the interval $[t_i, t_{i+1})$. Again, $M_i$ can only depend on the path of $S(t)$ up to time $t_i$.

Let $\mathcal{K}(t)$ denote the capital of Investor (expressed in cash) at time $t$. It is written as

$$\begin{aligned}&\mathcal{K}(0) = 1, \\ &\mathcal{K}(t) = \mathcal{K}(t_i) + M_i(S(t) - S(t_i)) \qquad \text{for } t_i \leq t < t_{i+1}.\end{aligned} \quad (1)$$

In the case $M_i < 0$, $\mathcal{K}(t)$ is the capital at time $t$ if he buys back $|M_i|$ units of the asset at the current price $S(t)$. As mentioned above, Investor is required to keep $\mathcal{K}(t)$ nonnegative, whatever price path $S(\cdot)$ Market chooses. Also, note that $\mathcal{K}(t)$ is continuous in $t$, since $S(t)$ is continuous in $t$.

By defining

$$\theta_i = \frac{M_i S(t_i)}{\mathcal{K}(t_i)},$$

we rewrite (1) as

$$\mathcal{K}(t) = \mathcal{K}(t_i)\left(1 + \theta_i \frac{S(t) - S(t_i)}{S(t_i)}\right) \qquad \text{for } t_i \leq t < t_{i+1}$$

in terms of the return $(S(t) - S(t_i))/S(t_i)$ of the asset.

In this paper, we mainly consider the scenario where Investor decides the trading times $t_1, t_2, \ldots$ by "limit order" strategy. Let $\delta_1, \delta_2 > 0$ be some constants and determine $t_1, t_2, \ldots$ as follows. After $t_i$ is determined, let $t_{i+1}$ be the first time after $t_i$ when either

$$\frac{S(t_{i+1})}{S(t_i)} = 1 + \delta_1 \quad \text{or} \quad \frac{S(t_{i+1})}{S(t_i)} = \frac{1}{1 + \delta_2}. \tag{2}$$

In this scheme, although Investor enters the market at time $t_0 = 0$, he begins trading at time $t_1$. This process leads to a discrete-time coin-tossing game embedded in the asset trading game as follows. Let

$$x_n = \frac{(1 + \delta_2)S(t_{n+1}) - S(t_n)}{(\delta_1 + \delta_2 + \delta_1\delta_2)S(t_n)} = \begin{cases} 1, & \text{if } S(t_{n+1}) = S(t_n)(1 + \delta_1), \\ 0, & \text{if } S(t_{n+1}) = S(t_n)/(1 + \delta_2). \end{cases}$$

The risk-neutral probability $\rho$ of the coin-tossing game [12, 17] can be deduced from

$$1 = \rho(1 + \delta_1) + \frac{1 - \rho}{1 + \delta_2},$$

which yields

$$\rho = \frac{\delta_2}{\delta_1 + \delta_2 + \delta_1\delta_2}.$$

Also, write

$$\tilde{\mathcal{K}}_n = \mathcal{K}(t_{n+1}), \qquad \nu_n = \frac{\delta_1 + \delta_2 + \delta_1\delta_2}{1 + \delta_2}\theta_n.$$

We then have the following protocol for an embedded discrete-time coin-tossing game.

EMBEDDED DISCRETE-TIME COIN-TOSSING GAME
**Protocol:**

$\tilde{\mathcal{K}}_0 := 1$.
FOR $n = 1, 2, \ldots$:
    Investor announces $\nu_n \in \mathbb{R}$.
    Market announces $x_n \in \{0, 1\}$.
    $\tilde{\mathcal{K}}_n = \tilde{\mathcal{K}}_{n-1}(1 + \nu_n(x_n - \rho))$.
END FOR

This embedded discrete-time game allows us to apply results on coin-tossing games to the asset trading game in continuous time. In particular, we can apply the strong law of large numbers for coin-tossing games.

However, it should be noted that in the embedded game, Market may decide to keep the variation of $S(t)$ small after $t_n$:

$$\frac{S(t_n)}{1 + \delta_2} < S(t) < S(t_n)(1 + \delta_1) \qquad \forall t \geq t_n.$$

The embedded coin-tossing game is then played for only $n$ rounds and the SLLN cannot be applied. Naturally, we are tempted to make $\delta_1, \delta_2$ smaller so that the total number of rounds increases, and we expect that Investor's high-frequency tradings take place when $\delta_1$ and $\delta_2$ are small. But once $\delta_1, \delta_2$ are announced, Market can always make the variation even smaller. This suggests that we should formulate the asset trading game and the move spaces of the players more carefully.

## 2.2. Formal definition of asset trading games and the notion of essential forcing

Here, we provide definitions of asset trading games and the move spaces of the players. Also, we define the notion of essential forcing of an event.

Market is required to choose a positive continuous function $S(\cdot)$ as his price path. Let

$$\Omega = C_{>0}[\mathbb{R}_+]$$

denote the set of positive continuous functions on $\mathbb{R}_+ = [0, \infty)$. This is the move space of Market, that is, Market chooses an element $S(\cdot) \in \Omega$. We also call $\Omega$ the *path space* or *sample space*. A subset $E$ of $\Omega$ is called an *event*. A *variable* is a real-valued function $f : \Omega \to \mathbb{R}$ on the path space.

In order to define the move space of Investor, we need a game-theoretic definition of a stopping time (see Section 5.3 of [15] and Section 1.1 of [8]) and a marked stopping time. A variable $\tau : \Omega \to [0, \infty]$ is called a *stopping time* if it follows from

$$\tau(S(\cdot)) < \infty \quad \text{and} \quad S(u) = \tilde{S}(u), \qquad 0 \leq u < \tau(S(\cdot))$$

that $\tau(\tilde{S}(\cdot)) = \tau(S(\cdot))$. Investor's trading times are stopping times. When $\tau(S(\cdot)) = t < \infty$, we say that $\tau$ is *realized* at time point $t$. Investor also decides how many units of the asset to hold at the time when $\tau$ is realized. A pair of variables

$$(\tau, m) : \Omega \to [0, \infty] \times \mathbb{R}$$

is a *marked stopping time* if $\tau$ is a stopping time and $m$ depends only on the path up to the realized time of $\tau$, that is,

$$\tau(S(\cdot)) < \infty \quad \text{and} \quad S(u) = \tilde{S}(u), \qquad 0 \le u < \tau(S(\cdot)),$$

implies that $m(\tilde{S}(\cdot)) = m(S(\cdot))$. We call $m$ the *mark* associated with the stopping time $\tau$. For definiteness, we define $m(S(\cdot)) = 0$ if $\tau(S(\cdot)) = \infty$.

A strategy $\mathcal{P}$ of Investor is a set of countably many marked stopping times

$$\mathcal{P} = \{(\tau_1, m_1), (\tau_2, m_2), \ldots\} \tag{3}$$

with the additional requirement that the stopping times are "discrete" in the following sense.

**Definition 2.1.** *A set of countably many stopping times $\{\tau_1, \tau_2, \ldots\}$ is discrete if for each $S(\cdot) \in \Omega$ there is no accumulation point of the set of realized stopping times.*

In the above definition, we are not requiring $\tau_1 \le \tau_2 \le \cdots$. For example, a strategy of Investor may consist of just two marked stopping times $\mathcal{P} = \{(\tau_1, m_1), (\tau_2, m_2)\}$, where $\tau_1$ is the first time $S(t)$ hits a predetermined high value and $\tau_2$ is the first time $S(t)$ hits a predetermined low value. Then, $\tau_1$ may realize before $\tau_2$ or vice versa. We use the notation $\tau_{(1)} \le \tau_{(2)} \le \cdots$ for the ordered realized stopping times.

By discreteness of the stopping times, we require Investor to trade only a finite number of times in every finite interval. The limit order type strategy in (2) clearly satisfies this requirement, because any continuous function on $[0, \infty)$ is uniformly continuous on the finite interval $[0, t]$. Under the above requirement, given a strategy $\mathcal{P}$ of Investor and a path $S(\cdot)$ of Market, the capital process $\mathcal{K}^{\mathcal{P}}(t) = \mathcal{K}^{\mathcal{P}}(t, S(\cdot))$ of Investor is defined as in (1), with $t_i = \tau_{(i)}(S(\cdot))$ and $M_i = m_{(i)}(S(\cdot))$, provided that the realized stopping times are all distinct. When realized time points of some stopping times coincide, for example when Investor employs nested strategies, we need to deal with obvious notational complications in adding up associated marks. But even when realized time points of some stopping times coincide, it is clear that the discreteness requirement guarantees that the capital process $\mathcal{K}^{\mathcal{P}}(t)$ is written as a finite sum for each $t > 0$.

Furthermore, we require that Investor observes his "collateral duty", that is, starting with the initial capital of $\mathcal{K}^{\mathcal{P}}(0) = 1$, his strategy $\mathcal{P}$ must satisfy

$$\mathcal{K}^{\mathcal{P}}(t, S(\cdot)) \ge 0 \qquad \forall t > 0, \forall S(\cdot) \in \Omega.$$

In summary, the move space $\mathcal{F}_0 = \{\mathcal{P}\}$ of Investor is the set of strategies in (3) satisfying the discreteness of Definition 2.1 and the collateral duty.

We note that $\mathcal{F}_0$ is closed under finite static mixtures. Let $\mathcal{P}_j = \{(\tau_{ij}, m_{ij})\}_{i=1}^{\infty}$, $j = 1, 2$, be two strategies belonging to $\mathcal{F}_0$. For $0 < c_1, c_2 < 1$ with $c_1 + c_2 = 1$, Investor sets up two accounts with the initial capitals $c_j$, $j=1,2$. He then applies $c_j \mathcal{P}_j = \{(\tau_{ij}, c_j m_{ij})\}_{i=1}^\infty$ to account $j$. This mixture is written as $c_1 \mathcal{P}_2 + c_2 \mathcal{P}_2 \in \mathcal{F}_0$ with the capital process $\mathcal{K}^{c_1 \mathcal{P}_1 + c_2 \mathcal{P}_2}(t) = c_1 \mathcal{K}^{\mathcal{P}_1}(t) + c_2 \mathcal{K}^{\mathcal{P}_2}(t)$. By induction, it is clear that $\mathcal{F}_0$ is closed with respect to any convex combination of a finite number of strategies.

In the spirit of game-theoretic probability, we assume that Investor first announces his strategy $\mathcal{P}$ to Market and then Market decides his path $S(\cdot)$. Therefore, the protocol of an asset trading game in continuous time is formulated as follows.

ASSET TRADING GAME IN CONTINUOUS TIME
**Protocol:**
  $\mathcal{K}(0) := 1$.
     Investor announces $\mathcal{P} \in \mathcal{F}_0$.
     Market announces $S(\cdot) \in \Omega$.

In game-theoretic probability, given some event $E \subset \Omega$, Investor is interpreted as the winner of the game if Market chooses a path $S(\cdot) \in E$ or else Investor's capital increases to infinity. In this case, we say that Investor can force the event $E$. In order to prove forcing of an event $E$, as shown in Shafer and Vovk [15], it is often useful to consider static mixtures of countably many strategies of Investor. However in our formulation, countable mixing involves a conceptual difficulty because such a mixture of trading strategies allows Investor to trade infinitely many times in a finite interval. Hence, in this paper, we use the following notion of "essential forcing" of an event $E$.

***Definition 2.2.*** *In the asset trading game in continuous time, Investor can essentially force an event $E$ if, for any $C > 0$, there exists a strategy $\mathcal{P}^C \in \mathcal{F}_0$ such that*

$$\sup_{0 \leq t < \infty} \mathcal{K}^{\mathcal{P}^C}(t, S(\cdot)) > C \quad \forall S(\cdot) \in E^c.$$

In Section 4, we will discuss the fact that essential forcing implies forcing in the sense of Shafer and Vovk [15] if we allow countable static mixtures. Therefore, the notion of essential forcing is good enough for the development in the present paper. Also, note that if Investor can essentially force a finite number of events $E_1, \ldots, E_K$, he can essentially force the intersection $E_1 \cap \cdots \cap E_K$ by a finite mixture of appropriate strategies (cf. Lemma 3.2 of [15]).

We also give a somewhat stronger definition of essential forcing for a finite interval $[T_1, T_2] \subset [0, \infty)$.

***Definition 2.3.*** *Investor can essentially force an event $E \subset \Omega$ in $[T_1, T_2]$ if, for any $C > 0$, there exists a strategy $\mathcal{P}^C \in \mathcal{F}$ such that*

$$\sup_{T_1 \leq t \leq T_2} \mathcal{K}^{\mathcal{P}^C}(t, S(\cdot)) > C \qquad \forall S(\cdot) \in E^c.$$

### 2.3. Variation exponent and Hölder continuity

Here, we summarize the notion of variation exponent and Hölder exponent (see, e.g., Section 4.1 of [6]). A continuous function $f$ on the interval $[T_1, T_2]$ is called Hölder continuous (Lipschitz continuous) of order $\overline{H}$ on $[T_1, T_2]$ if, for some $C > 0$,

$$\frac{|f(y) - f(x)|}{|y - x|^{\overline{H}}} \leq C, \qquad T_1 \leq \forall x < \forall y \leq T_2.$$

The largest value of such an $\overline{H}$ is usually called the *modulus of continuity* or the *Hölder exponent*. In this paper, we distinguish between several closely related notions and call $\overline{H}$ an *upper Hölder exponent*. In Section 3, we consider the set of functions

$$E_{\overline{H}, C, T_1, T_2} = \left\{ S \in \Omega \,\bigg|\, \frac{|\log S(y) - \log S(x)|}{|y - x|^{\overline{H}}} \leq C, T_1 \leq \forall x < \forall y \leq T_2 \right\}. \tag{4}$$

We also consider the bounding of the modulus of continuity (jaggedness of $S(\cdot)$) from below. Let $\mathbb{Q} \subset [0, \infty)$ be a given dense countable subset, such as the set of rational numbers. We define

$$\underline{E}_{\underline{H}, C, T_1, T_2} = \bigg\{ S \in \Omega \,\bigg|\, \forall \varepsilon > 0, \forall x \in [T_1, T_2 - \varepsilon] \cap \mathbb{Q}, \exists y \in (x, T_2]: \\ |\log S(y) - \log S(x)| \geq C\varepsilon^{\underline{H}} \text{ and } \frac{|\log S(y) - \log S(x)|}{|y - x|^{\underline{H}}} \geq C \bigg\}. \tag{5}$$

This definition of bounding the jaggedness from below by a *lower Hölder exponent* $\underline{H}$ is convenient for our limit order type strategy.

Finally, for $A > 0$, we write

$$E_{A, T_1, T_2} = \{ S \in \Omega \mid |\log S(y) - \log S(x)| \leq A, T_1 \leq \forall x < \forall y \leq T_2 \}. \tag{6}$$

The modulus of continuity can also be understood from the viewpoint of total variation of a continuous function. Here, we use the notion of strong $p$-variation from Section 11.6 of [15]. Let $\kappa: T_1 = t_0 < t_1 < \cdots < t_n = T_2$ be a division of the interval $[T_1, T_2]$. For $p \geq 1$ and a continuous function $f: [T_1, T_2] \to \mathbb{R}$, define

$$\overline{\mathrm{var}}_f(p) = \sup_{\kappa} \sum_{i=1}^{n} |f(t_i) - f(t_{i-1})|^p,$$

where sup is taken over all positive integers $n$ and all divisions $\kappa$. There exists a unique value $\overline{\mathrm{vex}}\, f \in [1, \infty]$ such that $\overline{\mathrm{var}}_f(p) < \infty$ for $p > \overline{\mathrm{vex}}\, f$ and $\overline{\mathrm{var}}_f(p) = \infty$ for $p < \overline{\mathrm{vex}}\, f$. We call $\overline{\mathrm{vex}}\, f$ the *variation exponent* of $f$. Note that each $S \in \Omega$ is uniformly continuous in the closed interval $[T_1, T_2]$ and hence bounded away from $0$ and $+\infty$. Also, $\log'(S) = 1/S$ is uniformly continuous in each compact interval of $(0, \infty)$. Therefore, $S \in E_{\overline{H}, C, T_1, T_2}$ implies $\overline{\mathrm{vex}} \log S \leq 1/\overline{H}$ and $S \in \underline{E}_{\underline{H}, C, T_1, T_2}$ implies $\overline{\mathrm{vex}} \log S \geq 1/\underline{H}$. Also, note that $\overline{\text{vex}} \log S = \overline{\text{vex}} S$ for $S \in \Omega$. From these relations, we call $H = 1/\overline{\text{vex}} S$ the *Hölder exponent* of $S$.

Results on the modulus of continuity of the paths of Brownian motion and fractional Brownian motion are summarized in Chapter IV of [1], Section 4.1 of [6] and Section 11.6 of [15].

### 2.4. Bayesian strategy for coin-tossing games

As discussed in Section 2.1, we mainly consider the scenario where Investor decides the trading times by the limit order type strategy in (2). In addition, we consider the situation where Investor specifies $M_i$ by the Bayesian strategy in [12]. Here, we briefly review the results of [12].

Suppose that Investor models Market's sequence of moves $x_1 x_2 \cdots$ ($x_i \in \{0, 1\}$) in the embedded discrete-time coin-tossing game of Section 2.1 by a probability distribution $Q$. Let $h_n = n\bar{x}_n = \sum_{i=1}^{n} x_i$ denote the number of heads and $t_n = n - h_n$ denote the number of tails. The beta-binomial model is defined as

$$Q(x_1 \cdots x_n) = \frac{1}{B(\alpha, \beta)} \int_0^1 p^{h_n + \alpha - 1} (1-p)^{t_n + \beta - 1} \, dp$$
$$= \frac{(\Gamma(\alpha + h_n)/\Gamma(\alpha)) \times (\Gamma(\beta + t_n)/\Gamma(\beta))}{\Gamma(\alpha + \beta + n)/\Gamma(\alpha + \beta)},$$

where $\alpha, \beta > 0$ are fixed and correspond to the prior numbers of heads and tails. We denote the conditional probability of $x_i = 1$ under $Q$ given $x_1, \ldots, x_{i-1}$ by

$$\hat{p}_i^Q = \hat{p}_i^Q(x_1, \ldots, x_{i-1}) = Q(x_i = 1 \mid x_1, \ldots, x_{i-1}).$$

In this model,

$$\hat{p}_n^Q = \frac{B(\alpha + h_{n-1} + 1, \beta + t_{n-1})}{B(\alpha + h_{n-1}, \beta + t_{n-1})} = \frac{\alpha + h_{n-1}}{\alpha + \beta + n - 1}$$

and the Investor's associated beta-binomial strategy is

$$\nu_n^* = \frac{\hat{p}_n^Q - \rho}{\rho(1-\rho)}. \tag{7}$$

The capital process $\tilde{\mathcal{K}}_n^*$ for this Bayesian strategy is explicitly written as

$$\tilde{\mathcal{K}}_n^*(x_1 \cdots x_n) = \frac{Q(x_1 \cdots x_n)}{\rho^{h_n}(1-\rho)^{t_n}}. \tag{8}$$

When both $h_n$ and $t_n$ are large, by using Stirling's formula

$$\log \Gamma(x) = (x - \tfrac{1}{2}) \log x - x + \log \sqrt{2\pi} + \mathrm{O}(x^{-1}),$$

we can evaluate the log capital process $\log \tilde{\mathcal{K}}_n^*$ as

$$\log \tilde{\mathcal{K}}_n^* = nD\left(\frac{h_n}{n} \bigg\| \rho\right) - \frac{1}{2}\log n + \mathrm{O}(1),$$

where

$$D(p\|q) = p\log\frac{p}{q} + (1-p)\log\frac{1-p}{1-q}$$

denotes the Kullback–Leibler information between $0 < p < 1$ and $0 < q < 1$. This expression, together with the Taylor expansion

$$D(\rho + \delta \| \rho) = \frac{\delta^2}{2\rho(1-\rho)} + \mathrm{O}(\delta^3),$$

allows us to analyze the behavior of the capital process for a high-frequency Bayesian strategy of Investor in the next section.

## 3. Essential forcing of variation exponent in the asset trading game

Consider the asset trading game in continuous time in Section 2.2 and the events $E_{\overline{H},C,T_1,T_2}$ in (4), $E_{\underline{H},C,T_1,T_2}$ in (5) and $E_{A,T_1,T_2}$ in (6). In this section, we prove the following main result of this paper.

**Theorem 3.1.** *For every $\overline{H} > 0.5, A > 0, C > 0, 0 \le T_1 < T_2 \le T$, Investor can essentially force*

$$E_{\overline{H},C,T_1,T_2} \Rightarrow E_{A,T_1,T_2}.$$

*For every $\underline{H} < 0.5, A > 0, C > 0, 0 \le T_1 < T_2 \le T$, Investor can essentially force*

$$E_{\underline{H},C,T_1,T_2} \Rightarrow E_{A,T_1,T_2}.$$

Here "$E_1 \Rightarrow E_2$" stands for $E_1^c \cup E_2$ (Section 4.1 of [15]) for two events $E_1, E_2 \subset \Omega$. Also, from the proof of the theorem below, it will be clear that Investor can essentially force these events in the interval $[T_1, T_2]$. This theorem roughly says that within an arbitrarily small constant $\varepsilon > 0$, Market's path is essentially forced to have the variation exponent $2 - \varepsilon < \overline{\mathrm{vex}}\, S < 2 + \varepsilon$, unless he stays constant. However, as we again discuss in Section 4, there is some gap between the two events (4) and (5). A stronger statement in terms of the variation exponent $\overline{\mathrm{vex}}\, S$ itself is now given in Theorem 1 of Vovk [21].

We give a proof of Theorem 3.1 after some preliminary investigations of the limit order type strategy in Section 2.1 combined with the Bayesian strategy in Section 2.4 for the embedded discrete-time game. Our proof is based on the limit order type strategy with sufficiently small $\delta_1 = \delta_2$ in (2). After the proof, we also investigate the behavior of Investor's capital processes for the cases where $\delta_1$ and $\delta_2$ decrease with different rates.

First, note that it suffices to consider the case $T_1 = 0$, because we can think of Investor as entering the game at time $t = T_1$ instead of $t = 0$ and using the strategy described below from $T_1$ on. Writing simply $T = T_2$, we thus consider only the case $[T_1, T_2] = [0, T]$.

We take the limit order type strategy in Section 2.1. Write $\delta = (\delta_1, \delta_2)$, where $\delta_1, \delta_2 > 0$. Let $t_0 = 0 < t_1 < t_2 < \cdots$ be the sequence of Investor's trading time points determined by (2). We then have the embedded discrete-time coin-tossing game and the associated $M_n$'s are determined by the Bayesian strategy in Section 2.4 in the form of $\nu_n^*$ in (7). The parameters $\alpha, \beta > 0$ for the Bayesian strategy are fixed throughout the rest of this section. It is clear that the resulting strategy $\mathcal{P} = \mathcal{P}^{\delta_1, \delta_2, \alpha, \beta}$ satisfies the collateral duty $\mathcal{K}^{\mathcal{P}}(t, S(\cdot)) \geq 0$, $\forall t > 0$, $\forall S(\cdot) \in \Omega$. We use the notation

$$\eta_i = \log(1 + \delta_i), \qquad \delta_i = e^{\eta_i} - 1, \qquad i = 1, 2,$$

and $\eta = (\eta_1, \eta_2)$. Define $n^* = n^*(T, \delta, S(\cdot))$ by $t_{n^*} < T \leq t_{n^*+1}$. Note that

$$n^*(T, \delta, S(\cdot)) \geq \frac{A}{\max(\eta_1, \eta_2)}$$

for every $S(\cdot) \in E_{A,0,T}^c$. Therefore, $n^*$ can be made arbitrarily large, uniformly in $S(\cdot) \in E_{A,0,T}^c$, by taking $\delta_1, \delta_2$ sufficiently small.

$\mathcal{K}(T) = \mathcal{K}^{\mathcal{P}^{\delta_1, \delta_2, \alpha, \beta}}(T, S(\cdot))$ is now written as

$$\mathcal{K}(T) = \tilde{\mathcal{K}}_{n^*}^* \left(1 + \theta_n^* \frac{S(T) - S(t_{n^*})}{S(t_{n^*})}\right), \qquad \theta_n^* = \frac{1 + \delta_2}{\delta_1 + \delta_2 + \delta_1 \delta_2} \nu_n^*.$$

Since $|\frac{S(T) - S(t_{n^*})}{S(t_{n^*})}| < \max(\delta_1, \delta_2)$, we have

$$\log \mathcal{K}(T) = \log \tilde{\mathcal{K}}_{n^*}^* + \mathrm{O}(1) = n^* D\left(\frac{h_{n^*}}{n^*} \Big\| \rho\right) - \frac{1}{2} \log n^* + \mathrm{O}(1). \tag{9}$$

Define

$$TV(\eta, T) = \sum_{i=1}^{n^*} |\log S(t_i) - \log S(t_{i-1})| = h_{n^*} \eta_1 + t_{n^*} \eta_2, \tag{10}$$

$$L(\eta, T) = \log S(t_{n^*}) - \log S(0) = h_{n^*} \eta_1 - t_{n^*} \eta_2, \tag{11}$$

$$\sigma(\eta, T) = \frac{L(\eta, T)}{TV(\eta, T)} = \frac{h_{n^*} \eta_1 - t_{n^*} \eta_2}{h_{n^*} \eta_1 + t_{n^*} \eta_2}.$$

We call $TV(\eta, T)$ the *total $\eta$-variation* of $\log S(t)$ in the interval $[0, T]$. We also write

$$L(T) = \log S(T) - \log S(0) = L(\eta, T) + \mathrm{O}(\max(\eta_1, \eta_2)).$$

We can then express (9) as

$$\log \mathcal{K}(T) = n^* D(p(\eta, T) \| \rho) - \tfrac{1}{2} \log n^* + \mathrm{O}(1), \tag{12}$$

where

$$p(\eta, T) = \frac{h_{n^*}}{n^*} = \frac{\eta_2(1 + \sigma(\eta, T))}{\eta_1(1 - \sigma(\eta, T)) + \eta_2(1 + \sigma(\eta, T))}.$$

Also, from (10) and (11), $n^*$ can be written as

$$n^* = \left( \frac{\eta_1 + \eta_2 - \sigma(\eta, T)(\eta_1 - \eta_2)}{2\eta_1 \eta_2} \right) TV(\eta, T). \tag{13}$$

Let $\eta_{1k} = a_1^{-k}, \eta_{2k} = a_2^{-k}$ for $a_1, a_2 > 1, k = 1, 2, \ldots$, and let $\log(1 + \delta_{1k}) = \eta_{1k}, \log(1 + \delta_{2k}) = \eta_{2k}$. We consider a sequence of the discretized games with $\delta_k = (\delta_{1k}, \delta_{2k})$ and let $\mathcal{K}_k(T)$ be the Investor's capital at $t = T$ for the beta-binomial strategy in each game. We denote the values of $n^*, \rho$ by $n_k^*, \rho_k$ corresponding to $\eta_k = (\eta_{1k}, \eta_{2k})$.

We are now ready to give a proof of Theorem 3.1.

**Proof of Theorem 3.1.** Take $a_1 = a_2 = a > 1$ and write $\eta_k = a^{-k}, \log(1 + \delta_k) = \eta_k$. We then have

$$n_k^* = \frac{TV(\eta_k, T)}{\eta_k}, \qquad p(\eta_k, T) = \frac{1 + \sigma(\eta_k, T)}{2}, \qquad \rho_k = \frac{1}{2 + \delta_k}.$$

Note that $\rho_k \to 1/2$ as $k \to \infty$. More precisely,

$$\rho_k = \frac{1}{2} - \frac{\delta_k}{4} + \mathrm{o}(\delta_k).$$

Consider $n_k^* D(p(\eta_k, T) \| \rho_k)$ in (12). Since $n_k^*$ can be made arbitrarily large uniformly in $S(\cdot) \in E_{A,0,T}^c$, we only need to consider $k$ and $S(\cdot) \in E_{A,0,T}^c$ such that $p(\eta_k, T)$ is close to $1/2$. Now use the Taylor expansion

$$D\left( \frac{1 + d_1}{2} \,\Big\|\, \frac{1 + d_2}{2} \right) = \frac{1}{2}(d_1 - d_2)^2 + \mathrm{o}(|d_1 - d_2|^2),$$

with $d_1 = \sigma(\eta_k, T), d_2 = -\delta_k/2$. Hence, noting $\delta_k = \mathrm{e}^{\eta_k} - 1 = a^{-k} + \mathrm{O}(a^{-2k})$, we can evaluate $n_k^* D(p(\eta_k, T) \| \rho_k)$ as

$$\begin{aligned}
n_k^* D(p(\eta_k, T) \| \rho_k) &\simeq a^k TV(\eta_k, T) \times \frac{1}{2} \left( \frac{L(T)}{TV(\eta_k, T)} + \frac{1}{2a^k} \right)^2 \\
&= \frac{1}{2} \left[ \frac{a^k}{TV(\eta_k, T)} L^2(T) + L(T) + \frac{1}{4} \frac{TV(\eta_k, T)}{a^k} \right].
\end{aligned} \tag{14}$$

Let $\overline{H} > 0.5$ and consider $S(\cdot) \in E_{\overline{H},C,T_1,T_2}$. It is easily seen that there exists some $c$ such that

$$TV(\eta_k, T) \leq ca^{Bk}, \qquad B = (1-\overline{H})/\overline{H} < 1$$

for all $k$ and for all $S(\cdot) \in E_{\overline{H},C,T_1,T_2}$. In this case, $a^k/TV(\eta_k,T) \to \infty$ as $k \to \infty$ uniformly in $S(\cdot) \in E_{\overline{H},C,T_1,T_2}$. As seen from the argument below at the end of the proof, for $S(\cdot) \in E_{A,0,T}^c$, we only need to consider the case $|L(T)| \geq A/4$. Therefore, $n_k^* D(p(\eta_k,T)\|\rho_k) \to \infty$ uniformly in $S(\cdot) \in E_{\overline{H},C,T_1,T_2}$. Also, it is easily verified that $\log n_k^*$ in (9) is of smaller order than $n_k^* D(p(\eta_k,T)\|\rho_k)$.

Now let $\underline{H} < 0.5$ and consider $S(\cdot) \in \underline{E}_{\underline{H},C,T_1,T_2}$. There then exist some $c$ and $k_0$ such that

$$TV(\eta_k, T) \geq ca^{Bk}, \qquad B = (1-\underline{H})/\underline{H} > 1$$

for all $k \geq k_0$ and all $S(\cdot) \in \underline{E}_{\underline{H},C,T_1,T_2}$. In this case, $TV(\eta_k,T)/a^k \to \infty$ as $k \to \infty$ uniformly in $S(\cdot) \in \underline{E}_{\underline{H},C,T_1,T_2}$. Again, $\log n_k^*$ can be ignored.

Thus, we have the following behavior of $\mathcal{K}_k(T)$ according to the values of the upper and the lower Hölder exponents.

If $\overline{H} > 0.5, S(\cdot) \in E_{\overline{H},C,T_1,T_2} \cap E_{A,0,T}^c$ and $|L(T)| \geq \dfrac{A}{4}$,  then $\mathcal{K}_k(T) \to \infty$.

If $\underline{H} < 0.5, S(\cdot) \in \underline{E}_{\underline{H},C,T_1,T_2} \cap E_{A,0,T}^c$, then $\mathcal{K}_k(T) \to \infty$.

We can guarantee the condition $|L(T)| \geq A/4$ above in the following manner. Let Investor divide his initial capital $\mathcal{K}(0) = 1$ into two accounts with the initial capitals $\mathcal{K}_1(0) + \mathcal{K}_2(0) = 1$. For the first account, Investor follows the high-frequency trading strategy explained above. For the second account, Investor starts the game at the first time $t_A(<T)$ when $|\log S(t_A) - \log S(0)| \geq A/2$ and follows the same high-frequency trading strategy. We denote Investor's capitals of respective accounts at $t = T$ by $\mathcal{K}_{k1}(T), \mathcal{K}_{k2}(T)$. Then,

$$\max(|\log S(T) - \log S(0)|, |\log S(T) - \log S(t_A)|) \geq \frac{A}{4}$$

on $E_{A,0,T}^c$. Therefore, at least one of $\mathcal{K}_{k1}(T), \mathcal{K}_{k2}(T)$ diverges to infinity. This proves the theorem. $\square$

For numerical comparison of capital processes, it is useful to approximate the capital process for the simple case. If $TV(\eta_k, T) \simeq ca^{Bk}$, then (14) is rewritten as

$$n_k^* D(p(\eta_k,T)\|\rho_k) \simeq \frac{1}{2}\left[\frac{a^{(1-B)k}}{c}L^2(T) + L(T) + \frac{ca^{(B-1)k}}{4}\right]. \qquad (15)$$

We also investigate the capital $\mathcal{K}_k(T)$ for two other cases: (ii) $a_1 < a_2$, (iii) $a_1 > a_2$. From (13) with $TV_k = TV(\eta_k, T), p_k = p(\eta_k, T)$, we have

$$n_k^* p_k \simeq \tfrac{1}{2}a_1^k(TV_k + L), \qquad n_k^*(1-p_k) \simeq \tfrac{1}{2}a_2^k(TV_k - L),$$

so it follows that

$$n_k^* D(p_k \| \rho_k) = n_k^* p_k \log \frac{p_k}{\rho_k} + n_k^*(1-p_k) \log \frac{1-p_k}{1-\rho_k}$$
$$\simeq \frac{1}{2}\left[ a_1^k(TV_k + L) \log \frac{p_k}{\rho_k} + a_2^k(TV_k - L) \log \frac{1-p_k}{1-\rho_k} \right]. \tag{16}$$

(ii) $a_1 < a_2$: In this case, $p_k, \rho_k \to 0$ as $k \to \infty$. However, the expression (16) has the following approximation:

$$n_k^* D(p_k \| \rho_k) \simeq \frac{1}{2} a_1^k(TV_k + L)\left[ \log \frac{TV_k + L}{TV_k - L} - \frac{2L}{TV_k + L} \right] \simeq \left( \frac{a_1^k}{TV(\eta_k, T)} \right) L^2(T). \tag{17}$$

Suppose that $TV(\eta_k, T) \simeq c a_1^{Bk}$. (17) is then rewritten as

$$n_k^* D(p_k \| \rho_k) \simeq \frac{a_1^{(1-B)k}}{c} L^2(T) \tag{18}$$

and we can derive the behavior of $\mathcal{K}_k(T)$ as follows:

$$\text{if } \overline{H} > 0.5 \text{ and } |L(T)| \geq \frac{A}{4}, \qquad \text{then } \mathcal{K}_k(T) \to \infty,$$

which is the only case such that $\mathcal{K}_k(T) \to \infty$.

(iii) $a_1 > a_2$: In this case, $p_k, \rho_k \to 1$ as $k \to \infty$. Again, the expression (16) has the following approximation:

$$n_k^* D(p_k \| \rho_k) \simeq \frac{1}{2} a_2^k(TV_k - L)\left[ \log \frac{TV_k - L}{TV_k + L} + \frac{2L}{TV_k - L} + \frac{2L}{TV_k - L} a_2^{-k} + \frac{1}{2} a_2^{-2k} \right]$$
$$\simeq \left( \frac{a_2^k}{TV(\eta_k, T)} \right) L^2(T) + L(T) + \frac{1}{4}\left( \frac{TV(\eta_k, T)}{a_2^k} \right). \tag{19}$$

Suppose that $TV(\eta_k, T) \simeq c a_2^{Bk}$. (19) is then rewritten as

$$n_k^* D(p_k \| \rho_k) \simeq \frac{a_2^{(1-B)k}}{c} L^2(T) + L(T) + \frac{c a_2^{(B-1)k}}{4} \tag{20}$$

and, as in the case of $a_1 = a_2$, the same behavior of $\mathcal{K}_k(T)$ is derived.

$$\text{If } \overline{H} > 0.5 \text{ and } |L(T)| \geq \frac{A}{4}, \qquad \text{then } \mathcal{K}_k(T) \to \infty.$$
$$\text{If } \underline{H} < 0.5, \qquad\qquad\qquad \text{then } \mathcal{K}_k(T) \to \infty.$$

We note that when $a = a_2$, the exponential growth part (20) is twice as large as (15).

## 4. Concluding remarks

In this paper, we proposed a new formulation of continuous-time games in the framework of the game-theoretic probability of Shafer and Vovk [15]. The present approach can be extended to prove that Investor can essentially force other properties of Market's path corresponding to various probability laws in continuous-time stochastic processes. Vovk [19] provided an approach to point processes and diffusion processes from the prequential viewpoint, but it was not developed further from the game-theoretic viewpoint. Extending the approach of the present paper, Vovk [20–22] has now fully developed the formulation of continuous-time processes in the game-theoretic probability setting within the conventional analysis.

From a theoretical perspective, it is important to consider taking the countable closure of the move space $\mathcal{F}_0$ of Investor. For the discrete-time games, there is no conceptual difficulty in considering static mixtures of countably many strategies. Even in continuous-time games, there is no conceptual difficulty in dividing the initial capital into countably many accounts and applying separate strategies to each account. Suppose that Investor can essentially force an event $E$. He can then divide his initial capital of one as

$$1 = \tfrac{1}{2} + \tfrac{1}{4} + \cdots$$

and put $1/2^i$ into the $i$th account as the initial capital. He applies the corresponding strategy $\mathcal{P}^C$ with $C = 2^i$ in Definition 2.2 to the $i$th account until $\mathcal{K}^{\mathcal{P}^{2^i}} \geq 1$. He then collects one (dollar) from each account and his capital diverges to infinity. This argument shows that if Investor can essentially force $E$, then he can force $E$, provided that static mixtures of countable strategies are allowed. In Vovk's new formulation [20–22], static mixtures of strategies are properly formulated and the notion of forcing is well defined for continuous-time games.

Our main Theorem 3.1 is stated in terms of the essential forcing of events (4) and (5). There is some gap between these two sets of functions. In particular, the set (5) may be too small. We used this definition for convenience in employing our simple limit order type strategy. A stronger statement in terms of the variation exponent $\overline{\text{vex}}\, S$ has been established in Theorem 1 of Vovk [21].

Finally, we comment on the differences between our approach and standard measure-theoretic approaches to no arbitrage and the Hölder exponent. A general theory of arbitrage was established by Delbaen and Schachermayer [4] (see also [5]) and it clarifies exact mathematical relations among various conditions concerning no arbitrage. Later, Rogers [14] gave an explicit trading strategy, but assumed a fractional Brownian motion with $H \neq 1/2$. In these measure-theoretic works, a stochastic process is given first and the effect of a trading strategy on probabilities over the set of paths is studied. On the contrary, we do not make any probabilistic assumptions. Furthermore, we study the behavior of an explicit strategy against each individual price path, rather than subsets of the set of paths.

## Acknowledgments

The authors are grateful to Norio Kono, Makoto Maejima and the referees for valuable comments and suggestions.